\documentstyle[psfig]{mn}

\begin{document}


\title{The scatter in the near-infrared colour-magnitude 
relation in spiral galaxies}

\author[R.F. Peletier and R.~de Grijs]{R.F.~Peletier$^{1,2}$ 
and R.~de Grijs$^{3,2}$\\
$^1$ Dept. of Physics, University of Durham, South Road, 
Durham DH1 3LE, UK\\
$^2$ Kapteyn Astronomical Institute, Postbus 800, 9700~AV~~Groningen, The Netherlands \\
$^3$ Astronomy Department, University of Virginia, PO Box 3818, 
Charlottesville, VA 22903-0818, USA}





\maketitle

\markboth{R.F.~Peletier and R.~de Grijs: The near-infrared
colour-magnitude relation}{}



\begin{abstract} 
We have determined a dust-free colour-magnitude (CM) relation for spiral 
galaxies, by using $I-K$ colours in edge-on galaxies above the plane.
We find that the scatter in this relation is small and approximately as large
as can be explained by observational uncertainties. The slope of the near-IR 
CM relation is steeper for spirals than for elliptical galaxies. We suggest
two possible explanations. First, the difference could be  caused by vertical
colour gradients in spiral galaxies. In that case these gradients should be
similar for all galaxies, on average $\sim$0.15 dex in  [Fe/H] per scale
height, and increase for later galaxy types.  The most likely explanation,
however, is
that spirals and ellipticals have intrinsically different CM relations. This
means that the stars in spirals are younger than those in ellipticals. The
age, however, or the fraction of young stars in spiral galaxies, would be
determined solely by the galaxy's luminosity, and not by its environment. 
\end{abstract}

\section{Introduction}

Baum (1959) and de Vaucouleurs (1961) first established that early-type
galaxies obey a tightly constrained colour -- absolute magnitude (CM)
relation. Somewhat later, in 1978,   Sandage \& Visvanathan first proposed
that the relation, if applied to galaxy clusters, could provide a significant
constraint on their past history of star formation. Using modern detectors
Bower et al. (1992a,b)  determined that both in Virgo and Coma the scatter
around the CM relations in $U-V$, $B-V$ and $V-K$ is extremely small ($\le$
0.05 mag), comparable to their observational uncertainties ($\sim$ 0.03 mag). 
This allowed them to
put strong constraints on the age scatter of early-type galaxies in clusters.
Recently, Ellis et al. (1997) again found very small scatter in the  CM
diagram of early-type galaxies in clusters at z $\sim$ 0.54, which led them to conclude that most of
the  star formation in these clusters must have ended at z $\sim$ 3. These
and other observations have led people to attribute the origin of the  CM
relation to changes in metallicity, caused by the fact that the larger
galaxies have larger binding energies, so that the gas can be enriched to
higher metallicities (e.g., Mathews \& Baker, 1971; Faber, 1977; Arimoto \&
Yoshii, 1987). 

For spiral galaxies, however, the situation is much more complicated, because
of recent star formation, as well as extinction by dust. Visvanathan \&
Griersmith (1977) found for early-type spiral galaxies in the Virgo cluster
(S0/a to Sab) within the errors the same optical CM relation as had been
found for E/S0 galaxies, except that the scatter was larger. Later Tully et
al. (1982) and  Mobasher et al. (1986) established optical-IR and IR CM
relations for early and late type spirals, with considerable scatter. Tully
et al. (1982) claimed that  the difference between early and late type spirals could
be explained by the presence of more star formation in late type spirals,
affecting  especially the blue light, and hence the $B-H$ colour. In the last
decade this picture has not changed very much. Interpreting a CM relation
using the $B-H$ colour is extremely complicated, given the fact that the
effects of star formation and extinction counteract each other.  Valentijn
(1990) and afterwards many others authors have shown that the extinction
corrections applied by Tully et al. are a factor of 2 or 3 too low, but in
any case very uncertain. To be able to draw more detailed conclusions about
galaxy formation and evolution, a different way of presenting the CM relation
has to be found. In this paper we argue that this can be done using the $I-K$
colour, obtained at positions  in the galaxy that are likely to be very
little affected by dust extinction. The main difference with previous work is
that we use local colours here, which is now possible due to advances in IR
detector technology. Integrated colours from, e.g., aperture photometry are
very hard to interpret due to the complicating effects of extinction and
stellar population differences,   except if the colours do not change much
inside the galaxy,  as is the case in elliptical galaxies. Secondly, to
eliminate the effects of very recent star formation as much as possible, we
have tried to use a colour that is as red as possible, while trying to keep
the wavelength baseline large. Knapen et al. (1995) show convincingly that
the $I-K$ colour is predominantly a dust/extinction indicator, while recent
star formation shows up much less than in a blue colour like $B-V$.

In this paper we try to use the $I-K$ vs. $M_K$ relation to determine a 
colour-magnitude relation with as low a scatter as possible for spiral
galaxies. This relation can be used to study the following problems: 1) The
nature  of spiral galaxies themselves. By comparing spirals with ellipticals 
we can study the star formation history of spirals in a very direct way,
making only the simple assumption that elliptical galaxies are coeval (see,
e.g., Kodama et al. 1998).  2) The role of the environment on the
evolution of galaxies. A tight CM relation for spirals can serve as a very
useful tool to study the evolution of spiral galaxies and the role of their
environment and 3) The use of the  CM diagram as a distance
indicator, very useful for  isolated galaxies in the field (de
Grijs \& Peletier, in preparation).

In the following section we present the data used for our CM relation. In
section 3 the scatter on the relation is discussed, and a comparison is
made  with early-type galaxies. In section 4 we discuss some implications of
the low scatter for the nature of spiral galaxies, after which we summarise
the paper.

\section{A dust-free CM relation}

To determine a dust-free CM diagram we decided to limit ourselves to two
datasets, of which we can be fairly confident that the amount of extinction
by  dust is very small. The first is a sample of 22 edge-on galaxies,
described in de Grijs et al. (1997). This is a random subsample of
Southern non-interacting galaxies with inclinations larger than 87$^{\rm o}$
and blue diameters D$^{B}_{25}$ larger than 2'.2. These galaxies were
observed in $I$ and $K$. More observational details can be found in
de Grijs (1998). It was found that although the central regions are
heavily obscured, the vertical colour profiles are symmetric, featureless and
with colour almost constant as a function of  radius (de Grijs et al. 1997). 
These 3 properties together  indicate that the colours here are  not reddened
by dust. We therefore took the average of the colour on both  sides of the
galaxy in the region where the colour profile is featureless. Absolute
$K$-magnitudes were determined using a simple Hubble flow model 
correcting for the motion of the local group (see de Grijs 1997). The second
sample is the sample of early-type spiral galaxies of Peletier \& Balcells
(1997). This is a sample of galaxies with inclinations larger than 50$^{\rm
o}$, for which the colours were determined in the bulge opposite to the
dustlane. The fact that the colour profiles on one side of the galaxy
generally are featureless shows that our assumption about negligible 
extinction is
justified.  However, since for some of the latest-type galaxies  of this
sample  the bulges are small, and therefore their colours probably more affected
by extinction, we only consider the earliest-type spirals (type $\le$ 1) for the
analysis done in this
paper.  Other samples are available (e.g., de Jong 1996), but less is known
about  their extinction properties,  and by including them the scatter
generally increases. 


The colours of both samples used were corrected for Galactic extinction and 
redshift in the way described in Balcells \& Peletier (1994),
Peletier \& Balcells (1997) and de Grijs (1997). Both corrections
are small (resp. $\le$ 0.10 and $\le$ 0.04 for $I-K$), because 
infrared colours of nearby galaxies are generally not very sensitive to these 
corrections. 

In order to compare these colours with central aperture measurements 
of ellipticals, a correction for internal colour gradients should be made.
The colours 
of the sample of Peletier \& Balcells (1997) were taken on the minor axis
in the bulge at 0.5 effective bulge radii. Peletier
\& Balcells (1996) and also Terndrup et al. (1994) show that
for their sample the colour difference in $I-K$ between the bulge 
and the inner disk (2 exponential scale lengths) is very small 
(0.07 $\pm$ 0.15 mag (Peletier \& Balcells 1996)), 
indicating small stellar population differences. 
For the edge-on galaxies of de Grijs et al. (1997) the colours were taken at an average 
height above the plane of 2 scale heights. 
Very little is known about vertical colour gradients in spirals.
In our Galaxy Trefzger et al. (1995) find a vertical metallicity gradient 
of $\Delta{\rm [Fe/H]}$/$\Delta~h_z$ = 0.05, using $h_z$ = 247 pc (Kent
et al. 1991). Using simple stellar population models (e.g. Vazdekis et al.
1996), this corresponds
to $\Delta(I-K)$/$\Delta~h_z$ = 0.025. For external galaxies at present no data 
is available for vertical stellar population gradients.
Radial colour gradients are small as well - 
for the sample of Peletier \& Balcells (1997) we find an average $R-K$
gradient of 0.12 mag per scale length, and a scatter of 0.11 mag. 
This corresponds to an $I-K$  gradient of about 0.10
mag/scale length. Some simple modelling shows that we need to
correct our colours by  $\le 0.075$ mag to make them compatible with those of 
Bower et al. (1992a). Since we do not know how vertical colour gradients
vary as a function of morphological type, we would have to apply the same
correction for each galaxy, and therefore the slope of the CM relation 
in spirals would not change. For this reason finally the correction 
for internal stellar population gradients was not applied.

\section{The scatter in the CM relation, and comparison with 
early-type galaxies}

\begin{figure}
\psfig{figure=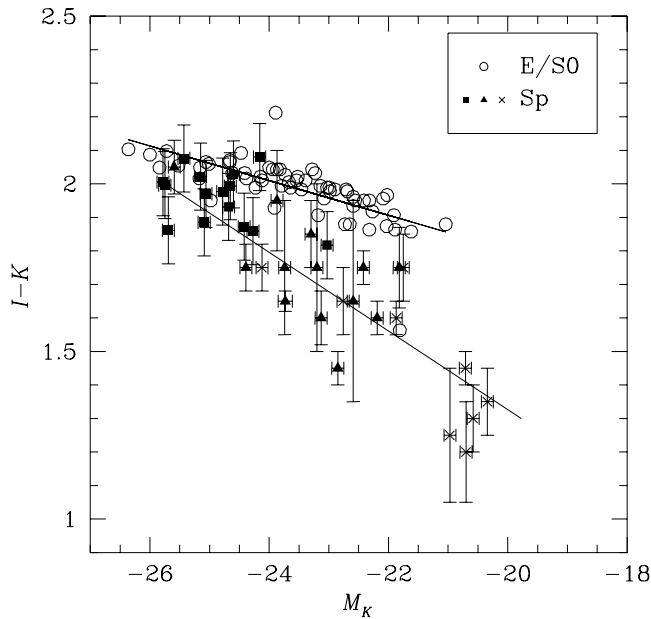,width=9cm}
\caption{The $I-K$ vs $M_K$ relation as discussed
in this paper. Plotted are the data from 
Peletier \& Balcells (1997) with type earlier than 1.5
(filled squares). The galaxies of de Grijs et al. (1997) are plotted as
filled triangles for types earlier than 5.5 and as crosses for later
types.  The elliptical/S0 galaxies from Bower et al. (1992a)
are shown as open circles.}
\end{figure}

In Fig.~1 we show the CM relation for our 2 samples. The drawn line is the 
least squares fit. We have also plotted the  elliptical and S0
galaxies from Bower  et al. (1992a), for which they found that the scatter
was comparable with the observational  uncertainties. Since these authors do
not have $I-K$  colours, we have converted their $V-K$ to $I-K$ using the
models of Vazdekis  et al. (1996). To do so we have made a linear fit of
$I-K$ as a function of $V-K$ for all single-age, single-metallicity models 
presented. Similar results are obtained if other stellar population models
(e.g., Worthey 1994) are used. Note that a sequence in morphological types
can be seen in Fig.~1, with the latest type spirals (filled circles) the
faintest and the bluest.

In Table~1 the best-fitting least squares fits are given. For 
our spiral galaxies we have applied a bivariate fitting routine, 
taking into account errors in both directions (see Peletier \& Willner 1993).
The uncertainties in our photometry are considerably larger than for the
ellipticals, because of  the uncertainties in calibrating the infrared
photometry with IRAC2b and the low light levels at large distances from
the galactic planes

\begin{table}
\caption[ ]{CM relations  ( $I-K$ = $a$ + $b$($M_K$ + 25) ) 
and scatter for ellipticals and spirals. (1)
gives the measured scatter in magnitudes, and (2) the scatter after
correcting for observational uncertainties. }
\begin{center}
\begin{tabular}{l|p{0.4cm}p{-0.1cm}c|p{1.1cm}cc|cc}
\hline
\hline
~ & \multicolumn{3}{|c|}{intercept} & \multicolumn{3}{c|}{slope} & (1) & (2) \\
\hline
E & 2.053 & $\pm$ & 0.008 &  --0.0438 & $\pm$ & 0.0041 & 0.037 & 0 \\
Sp     & 1.913 & $\pm$ & 0.030 & --0.1173  & $\pm$ & 0.0133 & 0.12  & 0.07 \\
\hline
\end{tabular}
\end{center}
\end{table}

\section{Discussion}

Although the scatter that we find in the CM relation for spiral galaxies is
much smaller than the values that have appeared in the literature until  now,
the results are consistent.  Tully  et al. (1982) already showed that spiral
galaxies obey  a CM relation, similar to the one found for ellipticals.  They
found that S0 galaxies occupy a completely different region on the ($B_T ~-~
H_{-0.5}$) CM plane than the later type spirals.  This result, however, is peculiar
for two reasons. First, some of the  S0 galaxies are redder than the bright
elliptical galaxies presented in the  same paper. Balcells \& Peletier (1994)
showed that bulges of early-type spirals and S0s in $U-R$, $B-R$ and $R-I$
are always bluer or have the same colour as ellipticals of the same
luminosity,  and therefore  these S0 galaxies must be affected considerably
by extinction  (at least A$_B$ = 0.5 mag), which seems to be in conflict with
the very small scatter among the S0 galaxies. Secondly, there is a
considerable gap of about 0.5 mag between the S0 galaxies and the early-type
spirals, also unlikely from other work. Similarly, the  relation of Wyse
(1982) for spirals is probably severely affected  by extinction, since her
brightest galaxies have $B-H$ values  around 5, much redder than the reddest
nearby ellipticals without dust. In any case, much better interpretable and
understandable is the work of Mobasher et al. (1986). They present CM
relations in infrared ($J-K$) and optical-infrared ($B-K$) colours for
spirals and ellipticals. In both bands the relation for spirals is steeper
than that for  ellipticals, although in the IR this difference is not very
significant.  Their result, although with large scatter, is largely
consistent with  our results presented here.

From Table~1 one can see that the scatter in the 
CM relation for spirals is very small (0.07 mag after taking
into account the observational errors, or possibly less, since this
number is smaller than the average observational error). 
Also, this relation falls about 0.1 -- 0.4 mag below the CM relation
for ellipticals, and has a steeper slope. Here we present two
possible explanations for the difference between the two CM relations.

\begin{enumerate}
\item While elliptical galaxies show very little or no star formation,
and have at most 10\%  of their stars formed in the last 5 Gyr (Bower  
et al. 1992b), all spiral galaxies are currently forming stars, whereby
the rate of star formation is determined {\it only} by the infrared
luminosity (or mass) of the galaxy. 
\item Spiral galaxies in general have considerable vertical colour
gradients. In this case we deduce that the vertical gradient per scale height 
needs to be about 0.15 dex in [Fe/H], if the dust-free CM relation 
is the same for
ellipticals and spirals. Later type spirals need to have larger gradients than
earlier types. 
\end{enumerate}

\begin{figure}
\psfig{figure=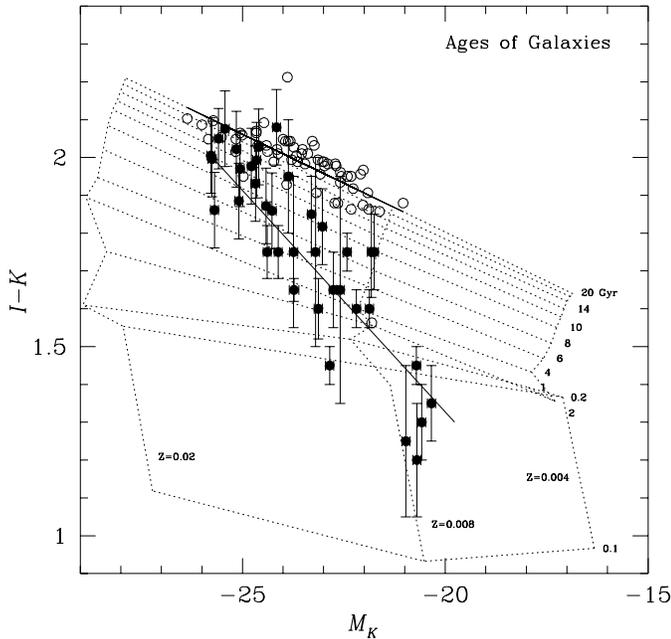,width=9cm}
\caption{The $I-K$ vs $M_K$ relation, now together with SSP models of
Bruzual \& Charlot (1998), for which metallicities and ages are given.}
\end{figure}

In the first case, the most straightforward explanation for the  origin of
both relations would be that they are driven by metallicity, but  that the
bluer $I-K$ colours are due to a small amount of recent star  formation, that
does not change the metallicity significantly, but affects the colours, and
somewhat brightens the galaxy. In Fig.~2 we show a grid of Single Stellar
Population (SSP) models of Bruzual \& Charlot (1998)  for metallicities between
0.004 and solar and ages between 0.1 and 20 Gyr. We have assumed that the
ellipticals are 20 Gyr old and lie on the CM relation, and have followed this
relation in time. Although the assumption that stars in a spiral galaxy are
coeval is almost certainly too simple, the figure shows that on average the
stars in late-type spirals are much younger than in ellipticals. The models
here show ages $\le$ 1 Gyr, although one can find other numbers if e.g.
models with exponentially decreasing star formation are used. The most
interesting aspect of this work is that we find that the  scatter in the CM
relation for spirals is very small, and that there is a gap between spirals and
ellipticals. This means that the current star formation  in a spiral galaxy
is determined by its size, morphological type or luminosity, and probably not by its
environment or interactions,  without much scatter. The CM relation can be
better understood by comparing it to the  large compilation of spiral
galaxies by McGaugh \& de Blok (1997), who show that fainter galaxies
generally have lower surface brightness, bluer optical colours, and
increasing gas mass fractions, as a consequence of which it is likely that
their average ages are also younger. 
 
Now, let us assume that  spiral galaxies  have the same $I-K$ CM relation as
ellipticals, but that the difference we observe here is solely caused by 
vertical stellar
population gradients in spiral galaxies. There are two reasons why we argue
that this option is not very likely. First, the vertical metallicity gradient
in our Galaxy, the only place where such a quantity has been measured at
present, would correspond to only 0.025 mag  per scale height in $I-K$, which
is by far not enough to explain the difference between spirals and
ellipticals. Secondly, the gradients would have to be larger for smaller
galaxies, whereas the scatter between the spirals of the same luminosity
would still need to remain very small. De Jong (1996) shows that radial colour
gradients for the later types are not larger than those for larger,
early-type spirals. For this reason it is not expected that the behavior in
the  vertical direction would be completely different.

\section{Conclusions}

The main results of this paper are as follows:

\begin{itemize}
\item We have determined a dust-free CM relation for spiral galaxies, by 
using $I-K$ colours in edge-on galaxies above the plane.
We find that the scatter in this relation is small and approximately as large as
can be explained by observational uncertainties. The slope of the IR 
CM-relation is steeper for spirals than for elliptical galaxies.
\item We present two possible explanations. First, the difference could be 
caused by vertical colour gradients in spiral galaxies. In that case these
gradients should be similar from galaxy to galaxy, have an average size
of about 0.15 dex in [Fe/H] per scale height, and increase for later galaxy
types. The other, much more likely, possibility is that spirals and 
ellipticals have 
different CM relations. The difference is caused by current star
formation, which has to be present in all spirals, as opposed to
ellipticals. The amount of 
current star formation would depend only on the galaxy's infrared 
luminosity, and not on its environment. 
\end{itemize}

\section*{Acknowledgements}
R.~de~Grijs thanks the Dept. of Physics of the University of Durham for
their hospitality during two visits. This paper is partly based on
observations collected at the Isaac Newton Telescope, La Palma,
the United Kingdom InfraRed Telescope, Mauna Kea, and the European
Southern Observatory, La Silla.

\end{document}